\documentclass[journal=apchd5,manuscript=letter,layout=twocolumn]{achemso}

\usepackage[version=3]{mhchem} 
\usepackage{amsmath}
\usepackage{amssymb}
\usepackage{amsfonts}
\usepackage{graphicx}
\usepackage{setspace}
\usepackage{tocloft}
\usepackage{xcolor}
\usepackage{bbm}
\usepackage{subfig}
\usepackage[countmax]{subfloat}
\usepackage{framed}
\usepackage{color}
\usepackage{hyperref}
\usepackage{epstopdf}
\usepackage{dsfont}
\usepackage{wrapfig}
\usepackage{relsize}
\usepackage{bm}
\usepackage{enumitem}
\usepackage{soul}
\usepackage{physics}

\author{Andrea Chiuri}
\email{andrea.chiuri@enea.it}
\phone{+39(06)94006049}
\affiliation[ENEA]
{ENEA - Centro Ricerche Frascati, Via E. Fermi 45, 00044, Frascati, Italy}
\author{Federico Angelini}
\affiliation[ENEA]
{ENEA - Centro Ricerche Frascati, Via E. Fermi 45, 00044, Frascati, Italy}
\author{Simone Santoro}
\affiliation[ENEA]
{ENEA - Centro Ricerche Frascati, Via E. Fermi 45, 00044, Frascati, Italy}
\author{Marco Barbieri}
\affiliation[RomaTre]
{Dipartimento di Scienze, Universit\`{a} degli Studi Roma Tre, Via della Vasca Navale 84, 00146, Rome, Italy}
\alsoaffiliation[CNR-INO]
{CNR - Istituto Nazionale di Ottica, Largo E. Fermi 6, 50125, Florence Italy}
\author{Ilaria Gianani}
\affiliation[RomaTre]
{Dipartimento di Scienze, Universit\`{a} degli Studi Roma Tre, Via della Vasca Navale 84, 00146, Rome, Italy}

\title{A Quantum Ghost Imaging Spectrometer}

\abbreviations{IR,NMR,UV}

\begin{document}

\begin{abstract}
We present a device that exploits spatial and spectral correlations in parametric downconversion at once. By using a ghost imaging arrangement, we have been able to reconstruct remotely the frequency profile of a composite system. The presence of distinct spectral regions is corroborated by a model-independent statistical analysis that constitutes an intriguing possibility also in the low count regime.
\end{abstract}

\section{}

Adopting correlations of light, either quantum or classical in nature~\cite{PhysRevA.52.R3429,PhysRevLett.87.123602,PhysRevLett.93.093602, PhysRevLett.96.063602, bennink04prl,Valencia05,bondani}, has allowed to develop novel solutions in sensing. The typical workhorse is a photon pair source, realised by parametric nonlinear optical effects~\cite{Shapiro15,padgett17ptrsa}. These processes convert individual photons from an intense pump light into photon pairs: as the elementary event must obey momentum and energy conservation, the spatial and spectral properties of one photon are entangled with the ones of the second photon. Notably, the two emissions can be in vastly different spectral regions\cite{Chan09,Aspden15,Kalachev_2008,Danino81,Borodin17}.

Thanks to the adoption of ghost imaging schemes, based on space and momentum correlations, images could be collected remotely, reducing the complexity of the detection after the object to a bucket detector with no resolution ability \cite{padgett17ptrsa}; the actual imaging system is placed on a correlated beam, and can be more conveniently located. The main appeal of the scheme is thus in its capability of displacing cumbersome analysis apparatuses to more convenient locations, when it comes to hardly accessible objects as well as frequency ranges. The analogue effect in the spectral domain has been termed quantum ghost spectroscopy, and relied on a conceptually identical scheme \cite{Yabushita04,Scarcelli,PhysRevX.4.011049}. Ghost techniques have demonstrated advantages for microscopy applications \cite{degiovanni} in terms of photon flux~\cite{Morris,Aspden15}, contrast~\cite{PhysRevA.72.013810}, and metrological performance \cite{Polino20,Chiuri22}.
Besides the two main spatial and spectral axes, ghost schemes have been extended to other degrees of freedom, foremost time and polarisation~\cite{Ryczkowski16,Chirkin18,PhysRevA.106.062601,Wu19,Leach10,Magnitskiy20}. 

More recently, the exploitation of induced coherence without induced emission~\cite{PhysRevA.44.4614,PhysRevA.100.053839} has allowed a similar advantage, by virtue of the same parametric processes, now differently used \cite{lemos14nat, Kalashnikov16,Paterova20,Kviatkovsky20,BarretoLemos:22}. The scheme uses a two-source arrangement, with the sample placed in between. The focus here is more about shifting the detection from the infrared domain - dense with characteristic lines of molecules - to less demanding and more efficient visible-light systems for either imaging or spectroscopy.

Spatial and spectral effects, however, are present simultaneously~\cite{PhysRevLett.95.260501}, thus the same system lends itself to build an imaging spectrometer that record  spectral and spatial information from remote at once~\cite{Basset19}. A first step in this direction has been taken in multispectral quantum imaging~\cite{Zhang23}: spectral correlations are used in order to determine the frequency, while imaging is performed conventionally.

In this article, we demonstrate the simultaneous usage of spectral and spatial correlations to realise a ghost imaging spectrometer. We adopt as our test case different filters placed in distinct positions. For our proof-of-principle demonstration we adopt degenerate light, in order to curtail efficiency limitations, and a 1D spatial geometry. 

This investigation allowed us to showcase the potential of our equipment, and prompted us to investigate model-agnostic methods for the discrimination of regions characterised by distinct spectra. These are proven to be resilient to limited data collection. Our method demonstrates a resource efficient use of quantum light, serving as a template for even more functional quantum devices \cite{Genovese_2016,Basset19}.

\section{The experiment}

\begin{figure*}
    \centering
    \includegraphics[width=\textwidth]{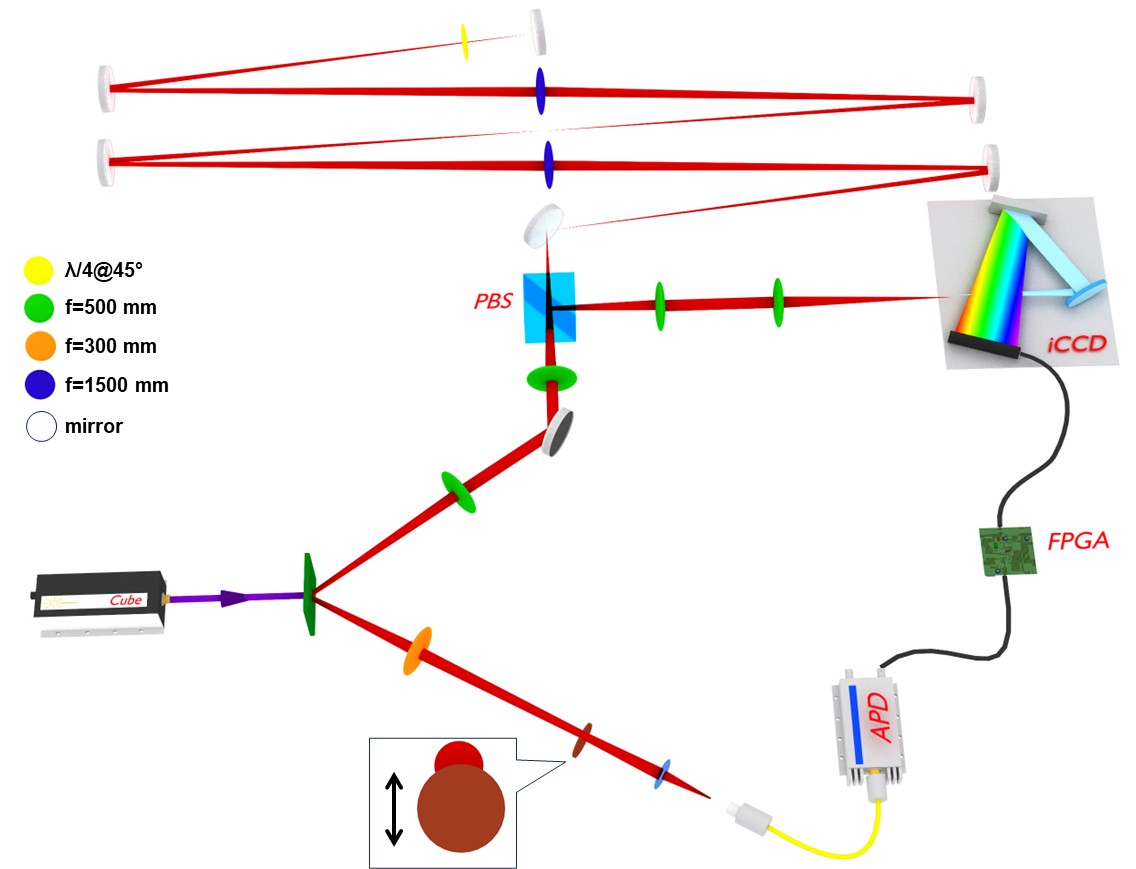}
    \caption{Experimental scheme. A CW laser at 405 nm arrives on a 3 mm-thick BBO crystal, generating signal-idler photon pairs by parametric down-conversion around the degenerate wavelength 810 nm. The idler is sent to the object  and then to the bucket detection; the object is a composite system of elements with different transmission and spectral response, studied only along its $y$ coordinate. The signal is sent through an image-preserving delay line, and then arrives onto a single-photon imaging spectrometer. The double-pass, image-preserving delay line accounts for the $75$ ns time needed to activate the ICCD from the APD trigger, hence its total length of $\approx 27 m$~\cite{Aspden15}. We used a polarizing beam splitter (PBS) to allow entry into the double-pass section of the delay line and then rotate the polarization by 90º within it line so that the photons can then exit at the PBS and reach the camera. A further telescope ensures that the beam is imaged on the entrance slit of the spectrometer. Finally, fine tuning of the time delay is achieved by means of a FPGA.}
    \label{Fig:setup}
\end{figure*}

We have implemented our ghost spectrometer with the setup shown in Fig.~\ref{Fig:setup}. 
The object to be analyzed, reached by the idler photon, is put at the image of the crystal plane - we thus exploit position correlations for ghost imaging. This is performed by a f=300 mm lens placed at 500 (750) mm from the crystal (target), resulting in a $1.5\times$ magnification of in the image of the BBO. The space- and frequency-insensitive detector (the 'bucket') is implemented by collecting photons in a multimode fibre through a high numerical aperture objective, and finally by an avalanche photodiode (APD).  Analysis is carried out on the correlated signal photon. This is performed by a spectrometer (Andor Kymera 328i) and an ICCD (Andor iStar DH334T-18U-73), preceded by an image-preserving delay line~\cite{Aspden13,Aspden15}. The system is able to provide a spatially resolved image over the vertical dimension $y$, while the orthogonal direction $x$ is devoted to the spectral analysis. The raw outcomes of our measurements are thus the number of counts from the ICCD for every pixel acquired in the photon-counting regime: we call each of such collections a $\lambda vsY$ map. 

\begin{figure*}
    \centering
    \includegraphics[width=\textwidth]{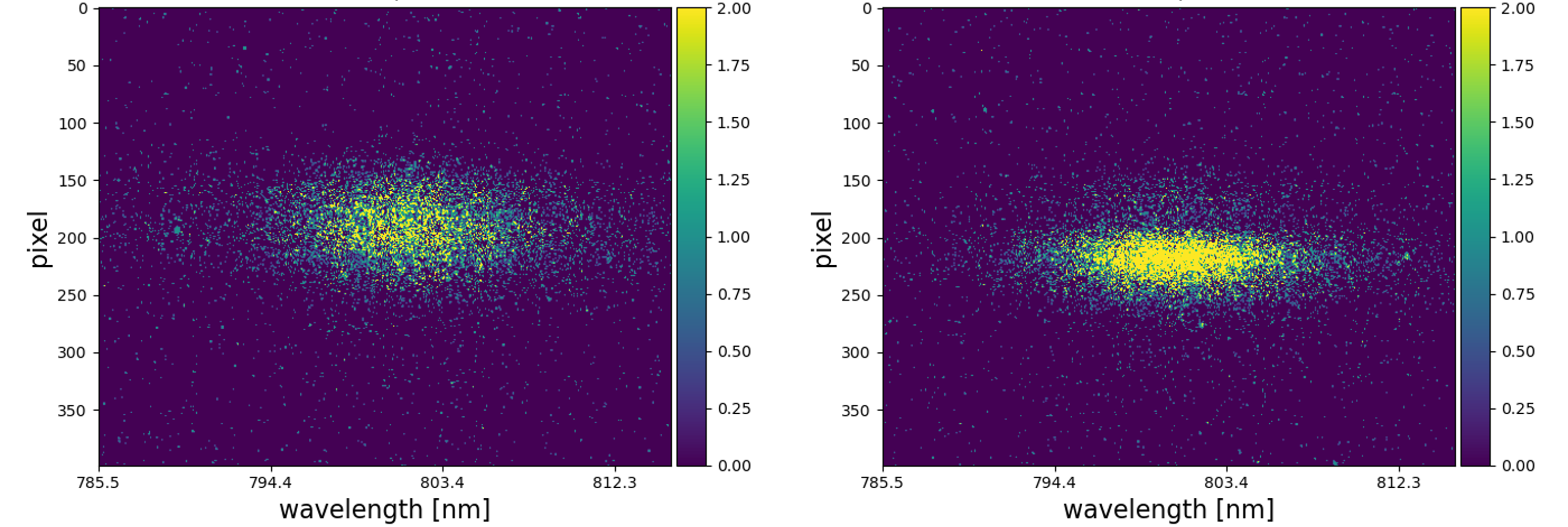}
    \caption{Imaging test of the setup. For the sake of obtaining a clearer presentation, the spectrum is limited by a filter placed immediately before the bucket detector, thus not acting as an object for imaging.  Left: reference measurement, without any object intercepting the beam in the bucket arm. Right: $\lambda vsY$ image retrieved when a neutral-density filter partially intercepts the beam in the bucket arm. A lower flux of photons was detected in the upper region, signalling the presence of the target object. In both panels the colour scale refers to the count rate in the acquisition time
    of 120 s (left) and 180 s (right).}   
    \label{neutral_filter}
\end{figure*}

The imaging capabilities of our setup are illustrated in Fig.~\ref{neutral_filter}, in which the left panels shows the $\lambda vsY$ map of the counts with no objects in the bucket arm. The right panel demonstrates how the image is modified when inserting a neutral-density filter only on part of the idler beam: its effect is shown as a reduction of the brightness in the upper part, with no significant dependence on the wavelength.  

\begin{figure*}
    \centering
    \includegraphics[width=\textwidth]{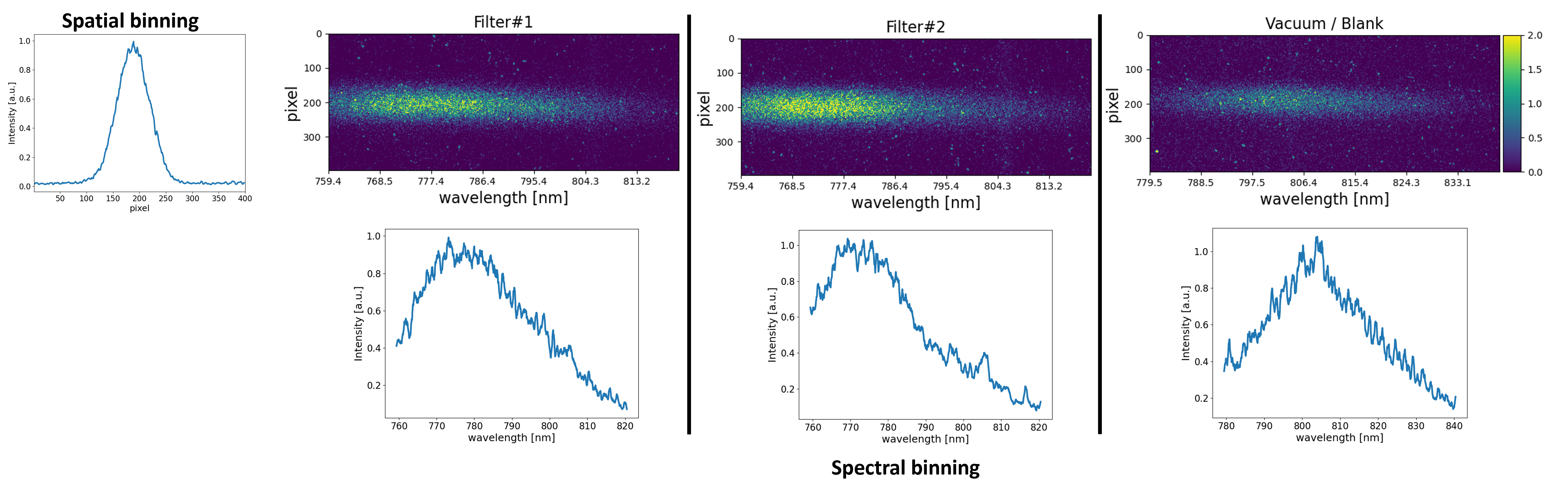}
 \caption{Reference spectra for the frequency-resolved measurements.  These maps are collected at long accumulation time (t=3600 s) in order to achieve a good signal to noise ratio. Since the filters are inserted completely on the idler beam, the spatial profiles are very similar in the three cases. The spectra are obtained by integrating over the spatial coordinate $y$, with their axis referring to the wavelength of the signal photon.}
    \label{fig:references}
\end{figure*}

\begin{figure*}
    \centering
    \includegraphics[width=\textwidth]{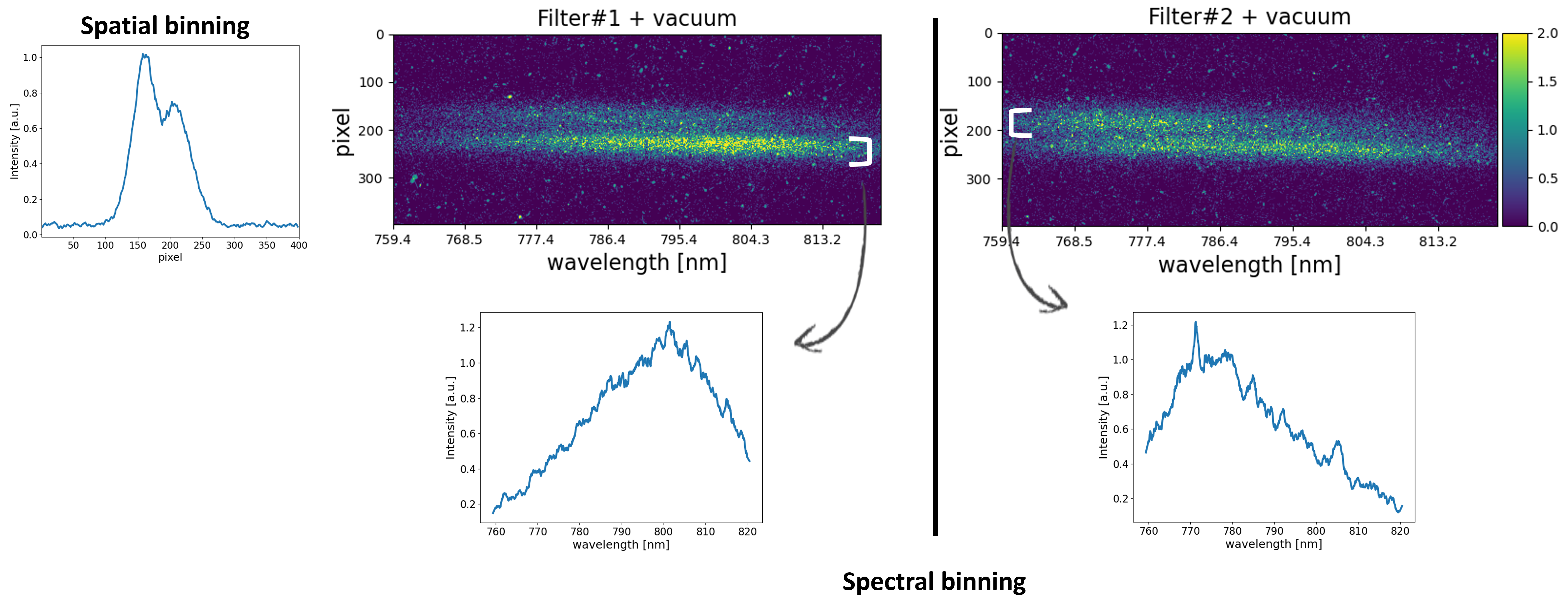}
    \caption{Frequency-resolved measurements with spatially separated spectral regions. The spatial binning alone gives indications on the presence of an object. The $\lambda vsY$ maps reveal the existence of distinct regions with disparate frequency reponses; the blank region correspond to the lower section of the map, as shown in the inset figure, while the filters correspond to the upper sections of the maps.}
    \label{fig:meas}
\end{figure*}

The joint spectral and spatial discrimination abilities of our setup can be assessed by introducing objects with different spectral responses, placed over distinct portions of the beam. In our work, we used two spectral filters with different transmission profiles. Their characterisation is presented in Fig.~\ref{fig:references}, in which we show three $\lambda vsY$ maps obtained when inserting either filter completely on the idler beam or leaving it unobstructed - this latter sets the blank reference. The marked difference among these three conditions is a clear indication of the spectral capabilities of our configuration. 
Figure \ref{fig:meas} reports the outcome of the measurements with the filters partially inserted on the idler beam: the simultaneous spectral analysis details the different frequency response from the two regions, with and without the filters. These can be limited by visual inspection, and the associated spectra are obtained by spatial integration over the relevant pixels.

\begin{figure*}
    \centering
    \includegraphics[width=\textwidth]{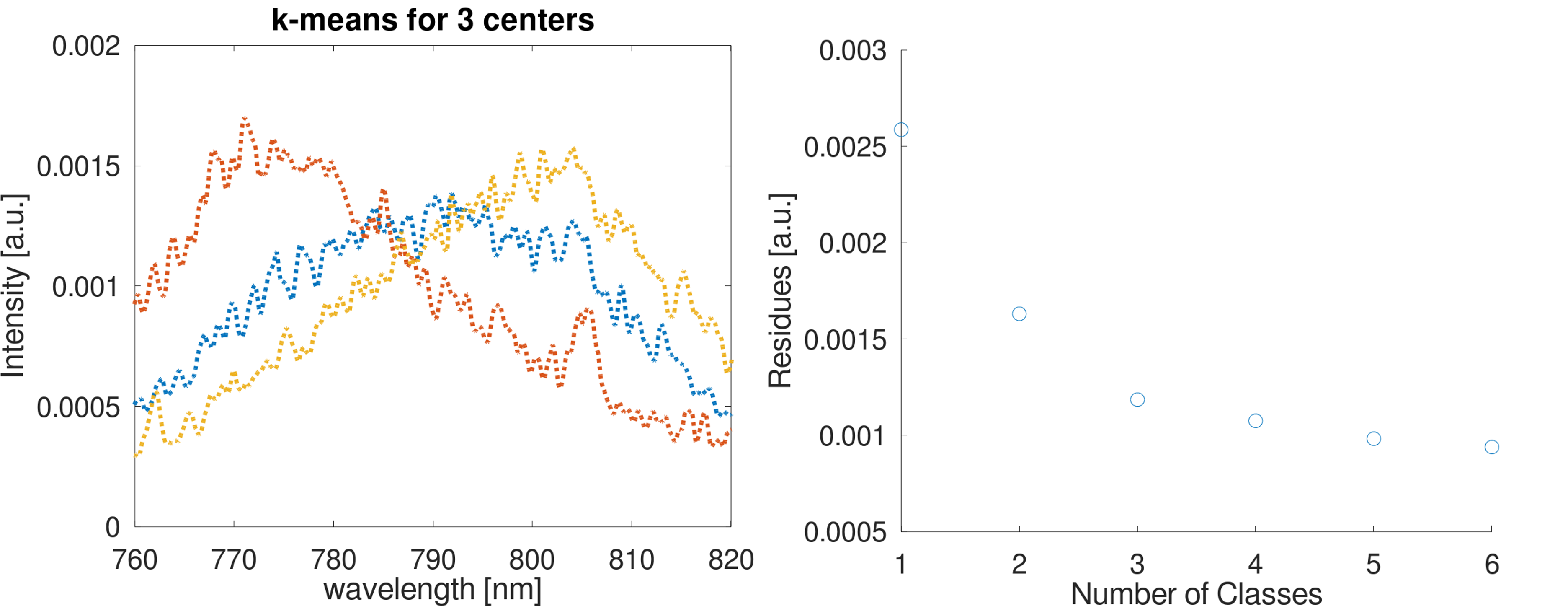}
    \caption{Results of the k means analysis. Left panel: estimated centroids relative to the three classes. Right panel: residual distance as a function of the number $k$ of clusters.}
    \label{fig:kmeans}
\end{figure*}

We illustrate a statistical method that identifies the three measured spectra, starting from the three reference spectra, as well as 5 repetitions of the measurements in Fig.~\ref{fig:meas} for each filter - each $\lambda vsY$ map thus yields two spectra, one per spatial-spectral region. In this way, we can account for the variablity in the acquisition of the spectra due to the uncertainties on the count rates. This considers the three following procedures:

\textbf{\emph{k means}}. The spectra are cast as vectors and k means analysis is applied in order to partition the dataset into $k$ clusters~\cite{}. This technique consists in using an iterative algorithm to group data around centroids acting as representatives of the classes. Optimisation is carried out as the minimisation of the average distance of the points to their closest centroid. In Fig.\ref{fig:kmeans} we show that 3 classes are sufficient to obtain good clustering of our dataset, since there is little difference in the residual distance for $k=3$ or for a larger number of classes. The three centroids can be easily related to the three reference spectra.  

\begin{figure*}
    \centering
    \includegraphics[width=\textwidth]{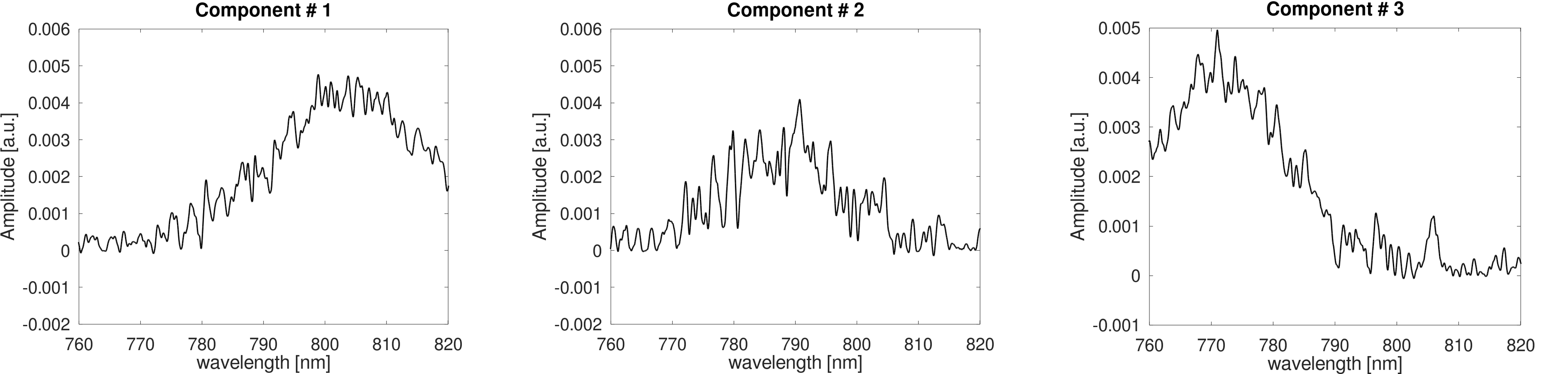}
    \caption{Components obtained via the NMF analysis. We can easily identify the first as the blank spectrum and the other two with the filtered spectra.}
    \label{fig:nmf}
\end{figure*}

\textbf{\emph{Non-negative Matrix Factorisation (NMF)}}
Any non-negative matrix $V$ can be decomposed as $V=WH+U$, where $W$ and $H$ are non-negative matrices, and $U$ is a matrix of residues, needed since the problem is approximated numerically in the typical instance. The matrix $H$ can be interpreted as that of the components vectors contributing to $V$, with $W$ the relative weights for each component~\cite{}. In our case, then, the matrix $V$ is given by the collection of our vectorised spectra. Based on the results of the k means analysis, we have fixed the number of components to three, leading to the spectra shown in Fig.~\ref{fig:nmf}. For each measured spectrum, a prominent component can be recognised by inspecting the corresponding column in the weight matrix $W$.

\begin{figure}
    \centering
    \includegraphics[width=\columnwidth]{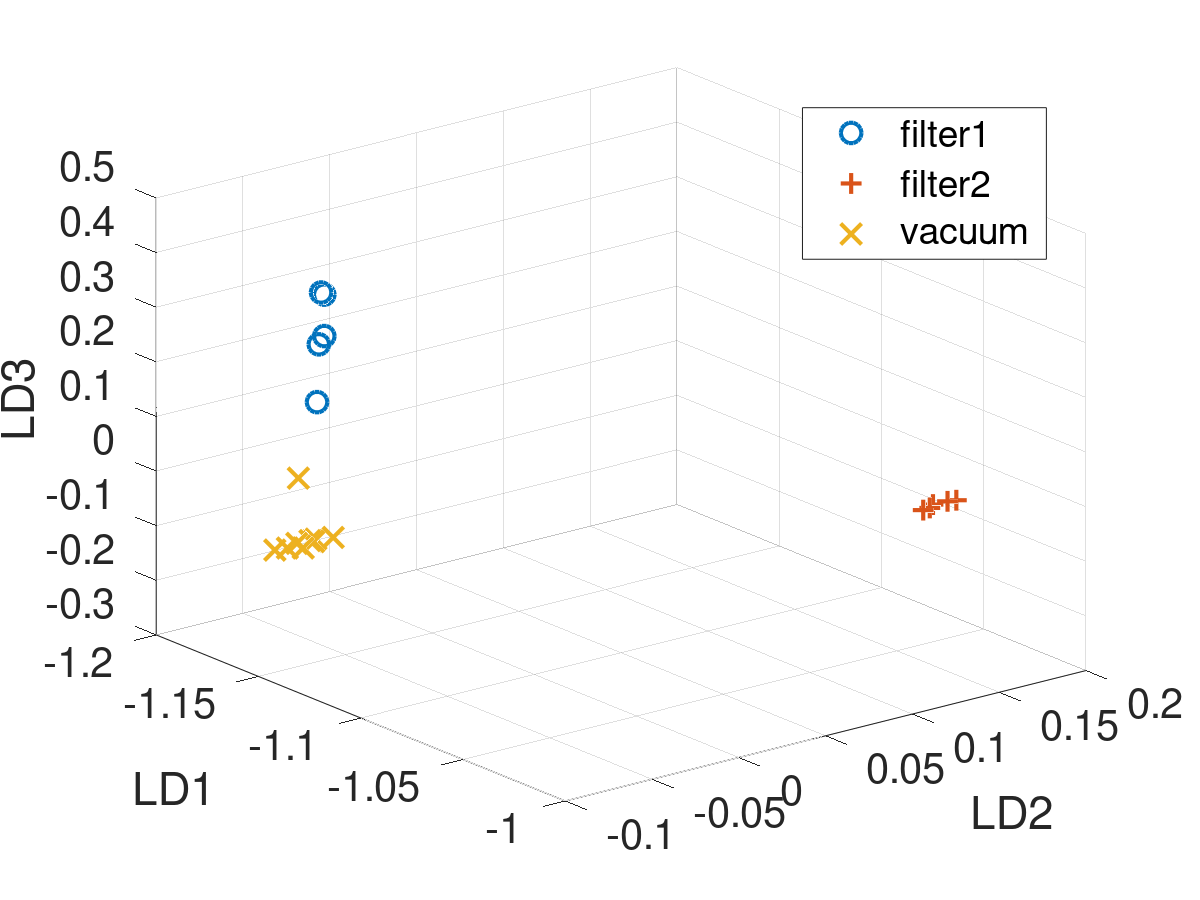}
    \caption{Results of LDA. Three Linear Discriminants (LD) are sufficient to identify the three classes in which the dataset can be divided.}
    \label{fig:lda}
\end{figure}

\textbf{\emph{Linear Discriminant Analysis (LDA)}}
LDA is a multivariate analysis combining the dimensionality reduction with data classification, in order to preserve as much information as possible on the discrimination of the different classes~\cite{}. The algorithm examines the directions along which the data have maximum variance and projects the data in this direction: \textit{noisy} directions are removed, achieving a representation of the data in lower dimension. The application of this algorithm to our dataset, presented in Fig.\ref{fig:lda}, reinforces once again the view that three components explain our observations, and that spectra can be sorted in three clusters.

\section{Conclusions}
The principles of a quantum ghost imaging spectrometer has been illustrated. Spatial and spectral degrees of freedom can be simultaneously harnessed and employed to obtain a spatially resolved spectra, combining frequency and position measurements. On the other hand, the presence of space-time coupling in the far field advises against adopting the scheme for momentum measurements.

We tested our setup with three spectral objects and the employment of several statistical methods allowed to support the conclusion that three classes can be identified. 

The specific experimental arrangement has been conceived and realised based on convenience of operation: the degenerate regime benefits from the availability of relatively high-efficiency detectors for both the bucket APD and the ICCD. 

The idea, however, is fully compatible with the employ of non-degenerate phase-matching, linking visible and infrared ranges in both space and frequency. 

\begin{acknowledgement}
The authors acknowledge the support of NATO thought the Science for Peace and Security (SPS) Program, project HADES (id G5839). 
\end{acknowledgement}

\bibliography{achemso-demo}

\providecommand{\latin}[1]{#1}
\makeatletter
\providecommand{\doi}
  {\begingroup\let\do\@makeother\dospecials
  \catcode`\{=1 \catcode`\}=2 \doi@aux}
\providecommand{\doi@aux}[1]{\endgroup\texttt{#1}}
\makeatother
\providecommand*\mcitethebibliography{\thebibliography}
\csname @ifundefined\endcsname{endmcitethebibliography}
  {\let\endmcitethebibliography\endthebibliography}{}
\begin{mcitethebibliography}{41}
\providecommand*\natexlab[1]{#1}
\providecommand*\mciteSetBstSublistMode[1]{}
\providecommand*\mciteSetBstMaxWidthForm[2]{}
\providecommand*\mciteBstWouldAddEndPuncttrue
  {\def\EndOfBibitem{\unskip.}}
\providecommand*\mciteBstWouldAddEndPunctfalse
  {\let\EndOfBibitem\relax}
\providecommand*\mciteSetBstMidEndSepPunct[3]{}
\providecommand*\mciteSetBstSublistLabelBeginEnd[3]{}
\providecommand*\EndOfBibitem{}
\mciteSetBstSublistMode{f}
\mciteSetBstMaxWidthForm{subitem}{(\alph{mcitesubitemcount})}
\mciteSetBstSublistLabelBeginEnd
  {\mcitemaxwidthsubitemform\space}
  {\relax}
  {\relax}

\bibitem[Pittman \latin{et~al.}(1995)Pittman, Shih, Strekalov, and
  Sergienko]{PhysRevA.52.R3429}
Pittman,~T.~B.; Shih,~Y.~H.; Strekalov,~D.~V.; Sergienko,~A.~V. Optical imaging
  by means of two-photon quantum entanglement. \emph{Phys. Rev. A}
  \textbf{1995}, \emph{52}, R3429--R3432\relax
\mciteBstWouldAddEndPuncttrue
\mciteSetBstMidEndSepPunct{\mcitedefaultmidpunct}
{\mcitedefaultendpunct}{\mcitedefaultseppunct}\relax
\EndOfBibitem
\bibitem[Abouraddy \latin{et~al.}(2001)Abouraddy, Saleh, Sergienko, and
  Teich]{PhysRevLett.87.123602}
Abouraddy,~A.~F.; Saleh,~B. E.~A.; Sergienko,~A.~V.; Teich,~M.~C. Role of
  Entanglement in Two-Photon Imaging. \emph{Phys. Rev. Lett.} \textbf{2001},
  \emph{87}, 123602\relax
\mciteBstWouldAddEndPuncttrue
\mciteSetBstMidEndSepPunct{\mcitedefaultmidpunct}
{\mcitedefaultendpunct}{\mcitedefaultseppunct}\relax
\EndOfBibitem
\bibitem[Gatti \latin{et~al.}(2004)Gatti, Brambilla, Bache, and
  Lugiato]{PhysRevLett.93.093602}
Gatti,~A.; Brambilla,~E.; Bache,~M.; Lugiato,~L.~A. Ghost Imaging with Thermal
  Light: Comparing Entanglement and ClassicalCorrelation. \emph{Phys. Rev.
  Lett.} \textbf{2004}, \emph{93}, 093602\relax
\mciteBstWouldAddEndPuncttrue
\mciteSetBstMidEndSepPunct{\mcitedefaultmidpunct}
{\mcitedefaultendpunct}{\mcitedefaultseppunct}\relax
\EndOfBibitem
\bibitem[Scarcelli \latin{et~al.}(2006)Scarcelli, Berardi, and
  Shih]{PhysRevLett.96.063602}
Scarcelli,~G.; Berardi,~V.; Shih,~Y. Can Two-Photon Correlation of Chaotic
  Light Be Considered as Correlation of Intensity Fluctuations? \emph{Phys.
  Rev. Lett.} \textbf{2006}, \emph{96}, 063602\relax
\mciteBstWouldAddEndPuncttrue
\mciteSetBstMidEndSepPunct{\mcitedefaultmidpunct}
{\mcitedefaultendpunct}{\mcitedefaultseppunct}\relax
\EndOfBibitem
\bibitem[Bennink \latin{et~al.}(2004)Bennink, Bentley, Boyd, and
  Howell]{bennink04prl}
Bennink,~R.~S.; Bentley,~S.~J.; Boyd,~R.~W.; Howell,~J.~C. Quantum and
  Classical Coincidence Imaging. \emph{Phys. Rev. Lett.} \textbf{2004},
  \emph{92}, 033601\relax
\mciteBstWouldAddEndPuncttrue
\mciteSetBstMidEndSepPunct{\mcitedefaultmidpunct}
{\mcitedefaultendpunct}{\mcitedefaultseppunct}\relax
\EndOfBibitem
\bibitem[Valencia \latin{et~al.}(2005)Valencia, Scarcelli, D'Angelo, and
  Shih]{Valencia05}
Valencia,~A.; Scarcelli,~G.; D'Angelo,~M.; Shih,~Y. Two-Photon Imaging with
  Thermal Light. \emph{Phys. Rev. Lett.} \textbf{2005}, \emph{94}, 063601\relax
\mciteBstWouldAddEndPuncttrue
\mciteSetBstMidEndSepPunct{\mcitedefaultmidpunct}
{\mcitedefaultendpunct}{\mcitedefaultseppunct}\relax
\EndOfBibitem
\bibitem[Bondani \latin{et~al.}(2012)Bondani, Allevi, and Andreoni]{bondani}
Bondani,~M.; Allevi,~A.; Andreoni,~A. Ghost imaging by intense multimode twin
  beam. \emph{The European Physical Journal Special Topics} \textbf{2012},
  \emph{203}, 151--161\relax
\mciteBstWouldAddEndPuncttrue
\mciteSetBstMidEndSepPunct{\mcitedefaultmidpunct}
{\mcitedefaultendpunct}{\mcitedefaultseppunct}\relax
\EndOfBibitem
\bibitem[Shapiro and Boyd(2012)Shapiro, and Boyd]{Shapiro15}
Shapiro,~J.; Boyd,~R. The physics of ghost imaging. \emph{Quantum Inf Process}
  \textbf{2012}, \emph{11}, 949--993\relax
\mciteBstWouldAddEndPuncttrue
\mciteSetBstMidEndSepPunct{\mcitedefaultmidpunct}
{\mcitedefaultendpunct}{\mcitedefaultseppunct}\relax
\EndOfBibitem
\bibitem[Padgett and Boyd(2017)Padgett, and Boyd]{padgett17ptrsa}
Padgett,~M.~J.; Boyd,~R.~W. An introduction to ghost imaging: quantum and
  classical. \emph{Philosophical Transactions of the Royal Society A:
  Mathematical, Physical and Engineering Sciences} \textbf{2017}, \emph{375},
  20160233\relax
\mciteBstWouldAddEndPuncttrue
\mciteSetBstMidEndSepPunct{\mcitedefaultmidpunct}
{\mcitedefaultendpunct}{\mcitedefaultseppunct}\relax
\EndOfBibitem
\bibitem[Chan \latin{et~al.}(2009)Chan, O'Sullivan, and Boyd]{Chan09}
Chan,~K. W.~C.; O'Sullivan,~M.~N.; Boyd,~R.~W. Two-color ghost imaging.
  \emph{Phys. Rev. A} \textbf{2009}, \emph{79}, 033808\relax
\mciteBstWouldAddEndPuncttrue
\mciteSetBstMidEndSepPunct{\mcitedefaultmidpunct}
{\mcitedefaultendpunct}{\mcitedefaultseppunct}\relax
\EndOfBibitem
\bibitem[Aspden \latin{et~al.}(2015)Aspden, Gemmell, Morris, Tasca, Mertens,
  Tanner, Kirkwood, Ruggeri, Tosi, Boyd, Buller, Hadfield, and
  Padgett]{Aspden15}
Aspden,~R.~S.; Gemmell,~N.~R.; Morris,~P.~A.; Tasca,~D.~S.; Mertens,~L.;
  Tanner,~M.~G.; Kirkwood,~R.~A.; Ruggeri,~A.; Tosi,~A.; Boyd,~R.~W.;
  Buller,~G.~S.; Hadfield,~R.~H.; Padgett,~M.~J. Photon-sparse microscopy:
  visible light imaging using infrared illumination. \emph{Optica}
  \textbf{2015}, \emph{2}, 1049--1052\relax
\mciteBstWouldAddEndPuncttrue
\mciteSetBstMidEndSepPunct{\mcitedefaultmidpunct}
{\mcitedefaultendpunct}{\mcitedefaultseppunct}\relax
\EndOfBibitem
\bibitem[Kalachev \latin{et~al.}(2008)Kalachev, Kalashnikov, Kalinkin,
  Mitrofanova, Shkalikov, and Samartsev]{Kalachev_2008}
Kalachev,~A.~A.; Kalashnikov,~D.~A.; Kalinkin,~A.~A.; Mitrofanova,~T.~G.;
  Shkalikov,~A.~V.; Samartsev,~V.~V. Biphoton spectroscopy in a strongly
  nondegenerate regime of SPDC. \emph{Laser Physics Letters} \textbf{2008},
  \emph{5}, 600\relax
\mciteBstWouldAddEndPuncttrue
\mciteSetBstMidEndSepPunct{\mcitedefaultmidpunct}
{\mcitedefaultendpunct}{\mcitedefaultseppunct}\relax
\EndOfBibitem
\bibitem[Danino and Freund(1981)Danino, and Freund]{Danino81}
Danino,~H.; Freund,~I. Parametric Down Conversion of X Rays into the Extreme
  Ultraviolet. \emph{Phys. Rev. Lett.} \textbf{1981}, \emph{46},
  1127--1130\relax
\mciteBstWouldAddEndPuncttrue
\mciteSetBstMidEndSepPunct{\mcitedefaultmidpunct}
{\mcitedefaultendpunct}{\mcitedefaultseppunct}\relax
\EndOfBibitem
\bibitem[Borodin \latin{et~al.}(2017)Borodin, Levy, and Shwartz]{Borodin17}
Borodin,~D.; Levy,~S.; Shwartz,~S. {High energy-resolution measurements of
  x-ray into ultraviolet parametric down-conversion with an x-ray tube source}.
  \emph{Applied Physics Letters} \textbf{2017}, \emph{110}, 131101\relax
\mciteBstWouldAddEndPuncttrue
\mciteSetBstMidEndSepPunct{\mcitedefaultmidpunct}
{\mcitedefaultendpunct}{\mcitedefaultseppunct}\relax
\EndOfBibitem
\bibitem[Yabushita and Kobayashi(2004)Yabushita, and Kobayashi]{Yabushita04}
Yabushita,~A.; Kobayashi,~T. Spectroscopy by frequency-entangled photon pairs.
  \emph{Phys. Rev. A} \textbf{2004}, \emph{69}, 013806\relax
\mciteBstWouldAddEndPuncttrue
\mciteSetBstMidEndSepPunct{\mcitedefaultmidpunct}
{\mcitedefaultendpunct}{\mcitedefaultseppunct}\relax
\EndOfBibitem
\bibitem[Scarcelli \latin{et~al.}(2003)Scarcelli, Valencia, Gompers, and
  Shih]{Scarcelli}
Scarcelli,~G.; Valencia,~A.; Gompers,~S.; Shih,~Y. {Remote spectral measurement
  using entangled photons}. \emph{Applied Physics Letters} \textbf{2003},
  \emph{83}, 5560--5562\relax
\mciteBstWouldAddEndPuncttrue
\mciteSetBstMidEndSepPunct{\mcitedefaultmidpunct}
{\mcitedefaultendpunct}{\mcitedefaultseppunct}\relax
\EndOfBibitem
\bibitem[Kalashnikov \latin{et~al.}(2014)Kalashnikov, Pan, Kuznetsov, and
  Krivitsky]{PhysRevX.4.011049}
Kalashnikov,~D.~A.; Pan,~Z.; Kuznetsov,~A.~I.; Krivitsky,~L.~A. Quantum
  Spectroscopy of Plasmonic Nanostructures. \emph{Phys. Rev. X} \textbf{2014},
  \emph{4}, 011049\relax
\mciteBstWouldAddEndPuncttrue
\mciteSetBstMidEndSepPunct{\mcitedefaultmidpunct}
{\mcitedefaultendpunct}{\mcitedefaultseppunct}\relax
\EndOfBibitem
\bibitem[Meda \latin{et~al.}(2015)Meda, Caprile, Avella, Ruo~Berchera,
  Degiovanni, Magni, and Genovese]{degiovanni}
Meda,~A.; Caprile,~A.; Avella,~A.; Ruo~Berchera,~I.; Degiovanni,~I.~P.;
  Magni,~A.; Genovese,~M. {Magneto-optical imaging technique for hostile
  environments: The ghost imaging approach}. \emph{Applied Physics Letters}
  \textbf{2015}, \emph{106}, 262405\relax
\mciteBstWouldAddEndPuncttrue
\mciteSetBstMidEndSepPunct{\mcitedefaultmidpunct}
{\mcitedefaultendpunct}{\mcitedefaultseppunct}\relax
\EndOfBibitem
\bibitem[Morris \latin{et~al.}(2015)Morris, Aspden, Bell, Boyd, and
  Padgett]{Morris}
Morris,~P.~A.; Aspden,~R.~S.; Bell,~J. E.~C.; Boyd,~R.~W.; Padgett,~M.~J.
  Imaging with a small number of photons. \emph{Nature Communications}
  \textbf{2015}, \emph{6}, 5913\relax
\mciteBstWouldAddEndPuncttrue
\mciteSetBstMidEndSepPunct{\mcitedefaultmidpunct}
{\mcitedefaultendpunct}{\mcitedefaultseppunct}\relax
\EndOfBibitem
\bibitem[D'Angelo \latin{et~al.}(2005)D'Angelo, Valencia, Rubin, and
  Shih]{PhysRevA.72.013810}
D'Angelo,~M.; Valencia,~A.; Rubin,~M.~H.; Shih,~Y. Resolution of quantum and
  classical ghost imaging. \emph{Phys. Rev. A} \textbf{2005}, \emph{72},
  013810\relax
\mciteBstWouldAddEndPuncttrue
\mciteSetBstMidEndSepPunct{\mcitedefaultmidpunct}
{\mcitedefaultendpunct}{\mcitedefaultseppunct}\relax
\EndOfBibitem
\bibitem[Polino \latin{et~al.}(2020)Polino, Valeri, Spagnolo, and
  Sciarrino]{Polino20}
Polino,~E.; Valeri,~M.; Spagnolo,~N.; Sciarrino,~F. {Photonic quantum
  metrology}. \emph{AVS Quantum Science} \textbf{2020}, \emph{2}, 024703\relax
\mciteBstWouldAddEndPuncttrue
\mciteSetBstMidEndSepPunct{\mcitedefaultmidpunct}
{\mcitedefaultendpunct}{\mcitedefaultseppunct}\relax
\EndOfBibitem
\bibitem[Chiuri \latin{et~al.}(2022)Chiuri, Gianani, Cimini, De~Dominicis,
  Genoni, and Barbieri]{Chiuri22}
Chiuri,~A.; Gianani,~I.; Cimini,~V.; De~Dominicis,~L.; Genoni,~M.~G.;
  Barbieri,~M. Ghost imaging as loss estimation: Quantum versus classical
  schemes. \emph{Phys. Rev. A} \textbf{2022}, \emph{105}, 013506\relax
\mciteBstWouldAddEndPuncttrue
\mciteSetBstMidEndSepPunct{\mcitedefaultmidpunct}
{\mcitedefaultendpunct}{\mcitedefaultseppunct}\relax
\EndOfBibitem
\bibitem[Ryczkowski \latin{et~al.}(2016)Ryczkowski, Barbier, Friberg, Dudley,
  and Genty]{Ryczkowski16}
Ryczkowski,~P.; Barbier,~M.; Friberg,~A.~T.; Dudley,~J.~M.; Genty,~G. Ghost
  imaging in the time domain. \emph{Nature Photonics} \textbf{2016}, \emph{10},
  167--170\relax
\mciteBstWouldAddEndPuncttrue
\mciteSetBstMidEndSepPunct{\mcitedefaultmidpunct}
{\mcitedefaultendpunct}{\mcitedefaultseppunct}\relax
\EndOfBibitem
\bibitem[Chirkin \latin{et~al.}(2018)Chirkin, Gostev, Agapov, and
  Magnitskiy]{Chirkin18}
Chirkin,~A.~S.; Gostev,~P.~P.; Agapov,~D.~P.; Magnitskiy,~S.~A. Ghost
  polarimetry: ghost imaging of polarization-sensitive objects. \emph{Laser
  Physics Letters} \textbf{2018}, \emph{15}, 115404\relax
\mciteBstWouldAddEndPuncttrue
\mciteSetBstMidEndSepPunct{\mcitedefaultmidpunct}
{\mcitedefaultendpunct}{\mcitedefaultseppunct}\relax
\EndOfBibitem
\bibitem[Restuccia \latin{et~al.}(2022)Restuccia, Gibson, Cronin, and
  Padgett]{PhysRevA.106.062601}
Restuccia,~S.; Gibson,~G.~M.; Cronin,~L.; Padgett,~M.~J. Measuring optical
  activity with unpolarized light: Ghost polarimetry. \emph{Phys. Rev. A}
  \textbf{2022}, \emph{106}, 062601\relax
\mciteBstWouldAddEndPuncttrue
\mciteSetBstMidEndSepPunct{\mcitedefaultmidpunct}
{\mcitedefaultendpunct}{\mcitedefaultseppunct}\relax
\EndOfBibitem
\bibitem[Wu \latin{et~al.}(2019)Wu, Ryczkowski, Friberg, Dudley, and
  Genty]{Wu19}
Wu,~H.; Ryczkowski,~P.; Friberg,~A.~T.; Dudley,~J.~M.; Genty,~G. Temporal ghost
  imaging using wavelength conversion and two-color detection. \emph{Optica}
  \textbf{2019}, \emph{6}, 902--906\relax
\mciteBstWouldAddEndPuncttrue
\mciteSetBstMidEndSepPunct{\mcitedefaultmidpunct}
{\mcitedefaultendpunct}{\mcitedefaultseppunct}\relax
\EndOfBibitem
\bibitem[Leach \latin{et~al.}(2010)Leach, Jack, Romero, Ireland, Franke-Arnold,
  Barnett, and Padgett]{Leach10}
Leach,~J.; Jack,~B.; Romero,~J.; Ireland,~D.; Franke-Arnold,~S.; Barnett,~S.;
  Padgett,~M. {Quantum imaging and orbital angular momentum}. Complex Light and
  Optical Forces IV. 2010; p 76130L\relax
\mciteBstWouldAddEndPuncttrue
\mciteSetBstMidEndSepPunct{\mcitedefaultmidpunct}
{\mcitedefaultendpunct}{\mcitedefaultseppunct}\relax
\EndOfBibitem
\bibitem[Magnitskiy \latin{et~al.}(2020)Magnitskiy, Agapov, and
  Chirkin]{Magnitskiy20}
Magnitskiy,~S.; Agapov,~D.; Chirkin,~A. Ghost polarimetry with unpolarized
  pseudo-thermal light. \emph{Opt. Lett.} \textbf{2020}, \emph{45},
  3641--3644\relax
\mciteBstWouldAddEndPuncttrue
\mciteSetBstMidEndSepPunct{\mcitedefaultmidpunct}
{\mcitedefaultendpunct}{\mcitedefaultseppunct}\relax
\EndOfBibitem
\bibitem[Wang \latin{et~al.}(1991)Wang, Zou, and Mandel]{PhysRevA.44.4614}
Wang,~L.~J.; Zou,~X.~Y.; Mandel,~L. Induced coherence without induced emission.
  \emph{Phys. Rev. A} \textbf{1991}, \emph{44}, 4614--4622\relax
\mciteBstWouldAddEndPuncttrue
\mciteSetBstMidEndSepPunct{\mcitedefaultmidpunct}
{\mcitedefaultendpunct}{\mcitedefaultseppunct}\relax
\EndOfBibitem
\bibitem[Lahiri \latin{et~al.}(2019)Lahiri, Hochrainer, Lapkiewicz, Lemos, and
  Zeilinger]{PhysRevA.100.053839}
Lahiri,~M.; Hochrainer,~A.; Lapkiewicz,~R.; Lemos,~G.~B.; Zeilinger,~A.
  Nonclassicality of induced coherence without induced emission. \emph{Phys.
  Rev. A} \textbf{2019}, \emph{100}, 053839\relax
\mciteBstWouldAddEndPuncttrue
\mciteSetBstMidEndSepPunct{\mcitedefaultmidpunct}
{\mcitedefaultendpunct}{\mcitedefaultseppunct}\relax
\EndOfBibitem
\bibitem[Lemos \latin{et~al.}(2014)Lemos, Borish, Cole, Ramelow, Lapkiewicz,
  and Zeilinger]{lemos14nat}
Lemos,~G.~B.; Borish,~V.; Cole,~G.~D.; Ramelow,~S.; Lapkiewicz,~R.;
  Zeilinger,~A. {Quantum imaging with undetected photons}. \emph{Nature}
  \textbf{2014}, \emph{512}, 409--412\relax
\mciteBstWouldAddEndPuncttrue
\mciteSetBstMidEndSepPunct{\mcitedefaultmidpunct}
{\mcitedefaultendpunct}{\mcitedefaultseppunct}\relax
\EndOfBibitem
\bibitem[Kalashnikov \latin{et~al.}(2016)Kalashnikov, Paterova, Kulik, and
  Krivitsky]{Kalashnikov16}
Kalashnikov,~D.; Paterova,~A.; Kulik,~S.; Krivitsky,~A.,~Leonid Infrared
  spectroscopy with visible light. \emph{Nature Photonics} \textbf{2016},
  \emph{10}, 98--101\relax
\mciteBstWouldAddEndPuncttrue
\mciteSetBstMidEndSepPunct{\mcitedefaultmidpunct}
{\mcitedefaultendpunct}{\mcitedefaultseppunct}\relax
\EndOfBibitem
\bibitem[Paterova \latin{et~al.}(2020)Paterova, Maniam, Yang, Grenci, and
  Krivitsky]{Paterova20}
Paterova,~V.,~Anna; Maniam,~M.,~Sivakumar; Yang,~H.; Grenci,~G.;
  Krivitsky,~A.,~Leonid Hyperspectral infrared microscopy with visible light.
  \emph{Science Advances} \textbf{2020}, \emph{6}, eabd0460\relax
\mciteBstWouldAddEndPuncttrue
\mciteSetBstMidEndSepPunct{\mcitedefaultmidpunct}
{\mcitedefaultendpunct}{\mcitedefaultseppunct}\relax
\EndOfBibitem
\bibitem[Kviatkovsky \latin{et~al.}(2020)Kviatkovsky, Chrzanowski, Avery,
  Bartolomaeus, and Ramelow]{Kviatkovsky20}
Kviatkovsky,~I.; Chrzanowski,~H.~M.; Avery,~E.~G.; Bartolomaeus,~H.;
  Ramelow,~S. Microscopy with undetected photons in the mid-infrared.
  \emph{Science Advances} \textbf{2020}, \emph{6}, eabd0264\relax
\mciteBstWouldAddEndPuncttrue
\mciteSetBstMidEndSepPunct{\mcitedefaultmidpunct}
{\mcitedefaultendpunct}{\mcitedefaultseppunct}\relax
\EndOfBibitem
\bibitem[Lemos \latin{et~al.}(2022)Lemos, Lahiri, Ramelow, Lapkiewicz, and
  Plick]{BarretoLemos:22}
Lemos,~G.~B.; Lahiri,~M.; Ramelow,~S.; Lapkiewicz,~R.; Plick,~W.~N. Quantum
  imaging and metrology with undetected photons: tutorial. \emph{J. Opt. Soc.
  Am. B} \textbf{2022}, \emph{39}, 2200--2228\relax
\mciteBstWouldAddEndPuncttrue
\mciteSetBstMidEndSepPunct{\mcitedefaultmidpunct}
{\mcitedefaultendpunct}{\mcitedefaultseppunct}\relax
\EndOfBibitem
\bibitem[Barreiro \latin{et~al.}(2005)Barreiro, Langford, Peters, and
  Kwiat]{PhysRevLett.95.260501}
Barreiro,~J.~T.; Langford,~N.~K.; Peters,~N.~A.; Kwiat,~P.~G. Generation of
  Hyperentangled Photon Pairs. \emph{Phys. Rev. Lett.} \textbf{2005},
  \emph{95}, 260501\relax
\mciteBstWouldAddEndPuncttrue
\mciteSetBstMidEndSepPunct{\mcitedefaultmidpunct}
{\mcitedefaultendpunct}{\mcitedefaultseppunct}\relax
\EndOfBibitem
\bibitem[Gilaberte~Basset \latin{et~al.}(2019)Gilaberte~Basset, Setzpfandt,
  Steinlechner, Beckert, Pertsch, and Gräfe]{Basset19}
Gilaberte~Basset,~M.; Setzpfandt,~F.; Steinlechner,~F.; Beckert,~E.;
  Pertsch,~T.; Gräfe,~M. Perspectives for Applications of Quantum Imaging.
  \emph{Laser \& Photonics Reviews} \textbf{2019}, \emph{13}, 1900097\relax
\mciteBstWouldAddEndPuncttrue
\mciteSetBstMidEndSepPunct{\mcitedefaultmidpunct}
{\mcitedefaultendpunct}{\mcitedefaultseppunct}\relax
\EndOfBibitem
\bibitem[Zhang \latin{et~al.}(2023)Zhang, England, and Sussman]{Zhang23}
Zhang,~Y.; England,~D.; Sussman,~B. Snapshot hyperspectral imaging with quantum
  correlated photons. \emph{Opt. Express} \textbf{2023}, \emph{31},
  2282--2291\relax
\mciteBstWouldAddEndPuncttrue
\mciteSetBstMidEndSepPunct{\mcitedefaultmidpunct}
{\mcitedefaultendpunct}{\mcitedefaultseppunct}\relax
\EndOfBibitem
\bibitem[Genovese(2016)]{Genovese_2016}
Genovese,~M. Real applications of quantum imaging. \emph{Journal of Optics}
  \textbf{2016}, \emph{18}, 073002\relax
\mciteBstWouldAddEndPuncttrue
\mciteSetBstMidEndSepPunct{\mcitedefaultmidpunct}
{\mcitedefaultendpunct}{\mcitedefaultseppunct}\relax
\EndOfBibitem
\bibitem[Aspden \latin{et~al.}(2013)Aspden, Tasca, Boyd, and Padgett]{Aspden13}
Aspden,~R.~S.; Tasca,~D.~S.; Boyd,~R.~W.; Padgett,~M.~J. EPR-based ghost
  imaging using a single-photon-sensitive camera. \emph{New Journal of Physics}
  \textbf{2013}, \emph{15}, 073032\relax
\mciteBstWouldAddEndPuncttrue
\mciteSetBstMidEndSepPunct{\mcitedefaultmidpunct}
{\mcitedefaultendpunct}{\mcitedefaultseppunct}\relax
\EndOfBibitem
\end{mcitethebibliography}

\end{document}